\documentclass[11pt]{article}
\usepackage[utf8]{inputenc}

\usepackage{geometry}                
\usepackage{graphicx}
\usepackage{amssymb}
\usepackage{amsthm}
\usepackage{amsmath}

\usepackage{natbib}
\usepackage[font=small]{caption}
\usepackage{dcolumn}

\newtheorem{prop}{Proposition}
\newtheorem{cor}{Corollary}

\usepackage{epstopdf}
\title{Conditional Value at Risk and Partial Moments for the Metalog Distributions}
\author{Valentyn Khokhlov}
\date{November 2019}

\begin{document}

\maketitle

\begin{abstract}
The metalog distributions represent a convenient way to approach many practical applications. Their distinctive feature is simple closed-form expressions for quantile functions. This paper contributes to further development of the metalog distributions by deriving the closed-form expressions for the Conditional Value at Risk, a risk measure that is closely related to the tail conditional expectations. It also addressed the derivation of the first-order partial moments and shows that they are convex with respect to the vector of the metalog distribution parameters.
\end{abstract}

\section{Introduction}

The metalog distribution family was developed by \cite{Keelin} and represents a system of continuous univariate probability distributions suitable for practical applications of fitting a distribution to the vector of known quantile values. As any metalog distribution quantile function has the closed-form expression that is linear with respect to the vector of the distribution parameters, it is easy to find those parameters when the set of quantile values is given (e.g. from the set of observations).

\cite{Keelin} defined the metalog distribution family using its quantile function as follows:
$$M_n(\alpha, \mathbf{a}) = \begin{cases}
a_{1}+a_{2}\ln \frac{\alpha }{1-\alpha } & \text{for }n=2, \\
M_2(\alpha, \mathbf{a})+a_{3}(\alpha-0.5)\ln \frac{\alpha }{1-\alpha } & \text{for }n=3, \\
M_3(\alpha, \mathbf{a})+a_{3}(\alpha-0.5) & \text{for }n=4, \\
M_{n-1}(\alpha, \mathbf{a})+a_{n}(\alpha-0.5)^{(n-1)/2} & \text{for odd }n \geq 5, \\
M_{n-1}(\alpha, \mathbf{a})+a_{n}(\alpha-0.5)^{(n/2-1)}\ln \frac{\alpha }{1-\alpha } & \text{for even }n \geq 6, \\

\end{cases}$$

The closed-form expression for some of the metalog distribution moments were derived in \cite{Keelin}. However, modern finance and risk management applications are largely based on the tail risk measures, especially after the coherent risk measures concept was introduced by \cite{Artzner}. One of the most popular coherent risk measure is the the Conditional Value at Risk (CVaR), also knows as the tail conditional expectation. In the paper we first focus on CVaR, and provide its closed-form expressions for the metalog distributions. After that, we derive the expressions for the first partial moments and show that they are convex with respect to the vector of the metalog distribution parameter. Their convexity allows applying a wide range of the optimization techniques.

In this paper we use basically the same mathematical notation and format of expressions as in \cite{Norton} in order to make the derived closed-form CVaR expression compatible with the comprehensive set of the CVaR formulas derived for many continuous probability distributions.

\section{Conditional Value at Risk for the Metalog Distribtuions}

Let the loss be represented by a real valued random variable $X$ that follows a metalog distribution with the quantile function $q_n(p, X) = M_n(p, \mathbf{a})$. The Conditional Value at Risk (CVaR), or the superquantile, at confidence level $\alpha$ is equal to the expected loss exceeding $M_n(\alpha, \mathbf{a})$, given by
$$ \bar{q}_n (\alpha, X) = E[ X | X > M_n(\alpha, \mathbf{a}) ] = \frac{1}{1-\alpha} \int_{M_n(\alpha, \mathbf{a})}^{\infty} xf_n(x, \mathbf{a})dx = \frac{1}{1-\alpha} \int_{\alpha}^{1} M_n(p, \mathbf{a}) dp$$
where $f_n(x, \mathbf{a})$ is the Probability Density Function (PDF) of $X$. Notice that we can represent the superquantile as an integral of the PDF or the quantile as $x = M_n(p, \mathbf{a})$. 

\begin{prop} \label{prop:CVaR}
Assume $X \sim Metalog_n(\mathbf{a})$, then its superquantile equals to
$$\bar{q}_n(\alpha, X) = \begin{cases}
{{a}_{1}}-{{a}_{2}}\ln \left[ \left( 1-\alpha  \right){{\alpha }^{\frac{\alpha }{1-\alpha }}} \right] & \text{for }n=2, \\
\bar{q}_2(\alpha, X) + \frac{a_3}{2}[\alpha ln\frac{\alpha}{1-\alpha}+1] & \text{for }n=3, \\
\bar{q}_3(\alpha, X) + \frac{a_4}{2}\alpha & \text{for }n=4, \\
\bar{q}_{n-1}(\alpha, X) + \frac{a_n}{1-\alpha}\frac{2}{n+1}\Big[0.5^\frac{n+1}{2}-(\alpha-0.5)^\frac{n+1}{2}\Big] & \text{for odd }n \geq 5, \\
\bar{q}_{n-1}(\alpha, X) + \frac{a_n}{1-\alpha}\frac{0.5^{k+1}}{k+1}\Big[\Psi\Big(1+\frac{k}{2}\Big)+\gamma+2ln2\Big] - \\ \qquad\qquad-\frac{a_n}{1-\alpha}\frac{(\alpha-0.5)^{k+1}}{(k+1)^{2}}\Big[{}_2F_1(1,k+1;k+2;1-2\alpha)-\\
\qquad\qquad-{}_2F_1(1,k+1;k+2;2\alpha-1)+(k+1)ln\frac{\alpha}{1-\alpha}\Big] & \text{for even }n \geq 6, \\

\end{cases}$$
where $\Psi$ is the digamma function, ${}_2F_1$ is the hypergeometric function, $\gamma$ is the Euler–Mascheroni constant, and for even $n \geq 6$, $k = \tfrac{n}{2} - 1$.

\begin{proof}
For $n=2$, 
\begin{align*}
    \bar{q}_2(\alpha, X) &= \frac{1}{1-\alpha}\int_{\alpha}^{1}{M_2(p,\mathbf{a})dp}=\frac{1}{1-\alpha}\int_{\alpha}^{1}{[a_1+a_2\ln\frac{p}{1-p}]dp}= \\
    & =\frac{{{a}_{1}}}{1-\alpha }\int_{\alpha }^{1}{dp}+\frac{{{a}_{2}}}{1-\alpha }\int_{\alpha }^{1}{\ln \frac{p}{1-p}dp} = \\
    & ={{a}_{1}}+\frac{{{a}_{2}}}{1-\alpha }\left[ \ln \left( 1-p \right)+p\ln \frac{p}{1-p}  \right]\Bigg|_{\alpha }^{1},
\end{align*}
and as $\lim\limits_{p \to 1^-} \left[\ln(1 - p) + p\ln\frac{p}{1 - p}\right] = 0$,
\begin{align*}
    \bar{q}_2(\alpha, X) &= {{a}_{1}}+\frac{{{a}_{2}}}{1-\alpha }\left[ -\ln \left( 1-\alpha  \right)-\alpha \ln \frac{\alpha }{1-\alpha }  \right]={{a}_{1}}-\frac{{{a}_{2}}}{1-\alpha }\ln \left[ {{\left( 1-\alpha  \right)}^{1-\alpha }}{{\alpha }^{\alpha }} \right]= \\ 
    & ={{a}_{1}}-\frac{{{a}_{2}}}{1-\alpha }\left[ \left( 1-\alpha  \right)\ln \left( 1-\alpha  \right)+\alpha \ln \alpha  \right]={{a}_{1}}-{{a}_{2}}\ln \left[ \left( 1-\alpha  \right){{\alpha }^{\frac{\alpha }{1-\alpha }}} \right]. 
\end{align*}
For $n=3$,
\begin{align*}
  \bar{q}_3(\alpha, X) &= \frac{1}{1-\alpha}\int_{\alpha}^{1}{M_3(p,\mathbf{a})dp}=\frac{1}{1-\alpha}\int_{\alpha}^{1}{\Big[M_2(p;\mathbf{a})+a_3\Big(p-\frac{1}{2}\Big)\ln\frac{p}{1-p}\Big]dp}= \\ 
  & = \bar{q}_2(\alpha, X) + \frac{a_3}{1-\alpha}\int_{\alpha }^{1}{p\ln\frac{p}{1-p}dp}-\frac{0.5a_3}{1-\alpha}\int_{\alpha }^{1}{\ln\frac{p}{1-p}dp}= \\ 
  & = \bar{q}_2(\alpha, X) + \frac{0.5{{a}_{3}}}{1-\alpha }\big[ \left( 1-\alpha  \right)\ln \left( 1-\alpha  \right)+\alpha \ln \alpha  \big]+\\
  &\qquad+ \frac{0.5{{a}_{3}}}{1-\alpha }\left[ -{{\alpha }^{2}}\ln \frac{\alpha }{1-\alpha }+\left( 1-\alpha  \right)-\ln \left( 1-\alpha  \right) \right]= \\ 
  & = \bar{q}_2(\alpha, X) + \frac{0.5a_3}{1-\alpha}\big[(1-\alpha)\ln (1-\alpha)+\alpha\ln\alpha-\alpha^2\ln\alpha+\alpha^2\ln(1-\alpha)+\\
  &\qquad+(1-\alpha)-\ln(1-\alpha)\big]= \\ 
  & = \bar{q}_2(\alpha, X) + \frac{0.5{{a}_{3}}}{1-\alpha }\big[ -\alpha \left( 1-\alpha  \right)\ln \left( 1-\alpha  \right)+\alpha \left( 1-\alpha  \right)\ln \alpha +\left( 1-\alpha  \right) \big]= \\ 
  & = \bar{q}_2(\alpha, X) + \frac{{{a}_{3}}}{2}\left[ \alpha \ln \frac{\alpha }{1-\alpha }+1 \right].
\end{align*}
For $n=4$,
\begin{align*}
  \bar{q}_4(\alpha, X) &= \frac{1}{1-\alpha }\int_{\alpha }^{1}{{{M}_{4}}\left( p;\mathbf{a} \right)dp}=\frac{1}{1-\alpha }\int_{\alpha }^{1}{\left[ {{M}_{3}}\left( p;\mathbf{a} \right)+{{a}_{4}}\left( p-\frac{1}{2} \right) \right]dp}= \\ 
  &= \bar{q}_3(\alpha, X) + \frac{{{a}_{4}}}{1-\alpha }\int_{\alpha }^{1}{pdp}-\frac{0.5{{a}_{4}}}{1-\alpha }\int_{\alpha }^{1}{dp}= \\ 
  &= \bar{q}_3(\alpha, X) + \frac{0.5{{a}_{4}}}{1-\alpha }\left( 1-{{\alpha }^{2}} \right)-\frac{0.5{{a}_{4}}}{1-\alpha }\left( 1-\alpha  \right)= \\ 
  &= \bar{q}_3(\alpha, X) + \frac{0.5{{a}_{4}}}{1-\alpha }\alpha \left( 1-\alpha  \right) = \bar{q}_3(\alpha, X) + \frac{a_4}{2}\alpha. 
\end{align*}
For odd $n \geq 5$ we can express $n = 2k+1$, $k = 2,3,...$, and
\begin{align*}
  \bar{q}_n(\alpha, X) &= \frac{1}{1-\alpha}\int_{\alpha}^{1}{{M_n}\left(p;\mathbf{a} \right)dp}=\frac{1}{1-\alpha}\int_{\alpha }^{1}{\left[ {M_{n-1}}\left(p;\mathbf{a} \right)+{a_n}{{\left( p-\frac{1}{2} \right)}^{k}} \right]dp}= \\ 
  &= \bar{q}_{n-1}(\alpha, X) + \frac{{{a}_{n}}}{1-\alpha }\int_{\alpha }^{1}{{{\left( p-\frac{1}{2} \right)}^{k}}dp}=\bar{q}_n(\alpha, X) + \frac{{{a}_{n}}}{1-\alpha }\frac{{{\left( p-0.5 \right)}^{k+1}}}{k+1}\Big|_{\alpha }^{1}= \\ 
  &= \bar{q}_{n-1}(\alpha, X) + \frac{{{a}_{n}}}{1-\alpha }\frac{2}{n+1}\left[ {{0.5}^{\frac{n+1}{2}}}-{{\left( \alpha -0.5 \right)}^{\frac{n+1}{2}}} \right]. 
\end{align*}
For even $n \geq 6$ we can express $n = 2k+2$, $k = 2,3,...$, and
$$
  \bar{q}_n(\alpha, X) = \frac{1}{1-\alpha }\int_{\alpha }^{1}{{{M}_{n}}\left( p;\mathbf{a} \right)dp}=\frac{1}{1-\alpha }\int_{\alpha }^{1}{\left[ {{M}_{n-1}}\left( p;\mathbf{a} \right)+{{a}_{n}}{{\left( p-\frac{1}{2} \right)}^{k}}\ln \frac{p}{1-p} \right]dp}.
$$  
Substituting $r = p - 0.5$ into the expression above,  
\begin{align*}
  \bar{q}_n(\alpha, X) &= \bar{q}_{n-1}(\alpha, X) + \frac{{{a}_{n}}}{1-\alpha } \int_{\alpha -0.5}^{0.5}{{{r}^{k}}\ln \frac{0.5+r}{0.5-r}dr}= \\ 
  &= \bar{q}_{n-1}(\alpha, X) + \frac{{{a}_{n}}}{1-\alpha }\frac{{{r}^{k+1}}}{{{\left( k+1 \right)}^{2}}}\Bigg[ {}_2F_1(1,k+1;k+2;-2r)-{}_2F_1(1,k+1;k+2;2r)+\\&+\left( k+1 \right)\ln \frac{1+2r}{1-2r} \Bigg]\Bigg|_{\alpha -0.5}^{0.5}.
 \end{align*}
For $r=0.5$ the sub-expression in the square brackets can be expanded using the digamma function as follows
\begin{align*}
  & \Bigg[ {}_2F_1(1,k+1;k+2;-2r)-{}_2F_1(1,k+1;k+2;2r)+\left( k+1 \right)\ln \frac{1+2r}{1-2r} \Bigg]\Bigg|_{r=0.5} = \\
  &= \frac{1+k}{2}\left( \Psi \left( \frac{2+k}{2} \right)-\Psi \left( \frac{1+k}{2} \right) \right)+\left( 1+k \right)\left( \gamma +\ln 2+\Psi \left( 1+k \right) \right).
\end{align*}
Considering that $\Psi \left( 1+k \right)=\frac{1}{2}\Psi \left( \frac{1+k}{2} \right)+\frac{1}{2}\Psi \left( \frac{2+k}{2} \right)+\ln 2$,
\begin{align*}
  & \frac{1+k}{2}\left( \Psi \left( \frac{2+k}{2} \right)-\Psi \left( \frac{1+k}{2} \right) \right)+\left( 1+k \right)\left( \gamma +\ln 2+\Psi \left( 1+k \right) \right) = \\ 
  & = \frac{1+k}{2}\left( \Psi \left( \frac{2+k}{2} \right)-\Psi \left( \frac{1+k}{2} \right) \right)+\left( 1+k \right)\left( \gamma +2\ln 2+\frac{1}{2}\Psi \left( \frac{1+k}{2} \right)+\frac{1}{2}\Psi \left( \frac{2+k}{2} \right)\right) = \\ 
  & =\frac{1+k}{2}\left[ 2\Psi \left( 1+\frac{k}{2} \right)+2\gamma +4\ln 2 \right]=\left( 1+k \right)\left[ \Psi \left( 1+\frac{k}{2} \right)+\gamma +2\ln 2 \right]. 
\end{align*}
Plugging this sub-expression into the formula for the superquantile,
\begin{align*} 
  \bar{q}_n(\alpha, X) &= \bar{q}_{n-1}(\alpha, X) + \frac{{{a}_{n}}}{1-\alpha }\frac{{{0.5}^{k+1}}}{k+1}\left[ \Psi \left( 1+\frac{k}{2} \right)+\gamma +2\ln 2 \right] - \frac{{{a}_{n}}}{1-\alpha }\frac{{{\left( \alpha -0.5 \right)}^{k+1}}}{{{\left( k+1 \right)}^{2}}} \times \\ 
  &\qquad\times \left[ {}_2F_1\left( 1,k+1;k+2;1-2\alpha  \right)-{}_2F_1\left( 1,k+1;k+2;2\alpha -1 \right)+\left( k+1 \right)\ln \frac{\alpha }{1-\alpha } \right]. 
\end{align*}
\end{proof}

\begin{cor} \label{cor:CVaR_6}
If $X \sim Metalog_6(\boldsymbol{a})$, its superquantile can be simplified to
$$\bar{q}_6(\alpha, X) = \bar{q}_5(\alpha, X) - \frac{a_6}{1-\alpha}\Bigg[\alpha\Bigg(\Big(\frac{\alpha^2}{3}-\frac{\alpha}{2}+\frac{1}{4}\Big)ln\frac{\alpha}{1-\alpha}+\frac{\alpha-1}{6}\Bigg)+\frac{ln(1-\alpha)}{12}\Bigg]$$
\end{cor}

\end{prop}

\section{First-order Partial Moments for the Metalog Distribution}

Assume that $X$ is a real valued random variable that follows a metalog distribution. Lets define $\alpha_w = Pr\{X \leq w\}$. By definition, $F_X(w) = \alpha_w$.

\begin{prop} \label{prop:part_mom1}
The first-order partial moments at w for a metalog distribution can be expressed as
$$\mu_1^{+}(w) = (1 - \alpha_w)(\bar{q}_n(\alpha_w, X) - w),$$
$$\mu_1^{-}(w) = w\alpha_w - E[X] + (1-\alpha_w)\bar{q}_n(\alpha_w, X).$$

\begin{proof}
The upper partial moment, by definition, $\mu_1^{+}(w) = \int_w^{+\infty}{(x-w)f(x)dx}$.
By substituting $p = F(x)$, $dp = f(x)dx$, $x = F^{-1}(p) = M_n(p, \mathbf{a})$, and considering that $F(+\infty) = 1$ and $F(w) = \alpha_w$,
\begin{align*}
  \mu_1^{+}(w) &= \int_w^{+\infty}{(x-w)f(x)dx} = \int_w^{+\infty}{xf(x)dx} - w\int_w^{+\infty}{f(x)dx} =\\
  &= \int_{\alpha_w}^1{M_n(p, \mathbf{a})dp} - w\int_{\alpha_w}^1{dp} = (1-\alpha_w)\bar{q}_n(\alpha_w,X) - w(1 - \alpha_w) =\\
  &= (1 - \alpha_w)(\bar{q}_n(\alpha_w, X) - w).
\end{align*}
The lower partial moment, by definition, $\mu_1^{-}(w) = \int_{-\infty}^{w}{(w-x)f(x)dx}$.
Performing the same substitutions,
\begin{align*}
  \mu_1^{-}(w) &= \int_{-\infty}^{w}{(w-x)f(x)dx} = w\int_{-\infty}^{w}{f(x)dx} - \int_{-\infty}^{w}{xf(x)dx} =\\
  &= w\int_{-\infty}^{w}{f(x)dx} - \left(\int_{-\infty}^{+\infty}{xf(x)dx} - \int_{w}^{+\infty}{xf(x)dx}\right) =\\
  &= w\int_0^{\alpha_w}{dp} - \left(E[X]-\int_{\alpha_w}^1{M_n(p, \mathbf{a})dp}\right) =\\
  &= w\alpha_w - E[X] + (1-\alpha_w)\bar{q}_n(\alpha_w, X).
\end{align*}
\end{proof}

\end{prop}

Lets define the upper quantile function $M_n^{+}(w; p, \mathbf{a}) = max\{M_n(p, \mathbf{a}) - w, 0\}$ and the lower quantile function $M_n^{-}(w; p, \mathbf{a}) = max\{w - M_n(p, \mathbf{a}), 0\}$. As the quantile function is non-decreasing, $M_n(p,\mathbf{a}) \leq w\enspace\forall p < \alpha_w$ and $M_n(p,\mathbf{a}) \geq w\enspace\forall p \geq \alpha_w$. 

\begin{prop} \label{prop:part_mom2}
The first-order partial moments at w for a metalog distribution can be expressed as
$$\mu_1^{+}(w) = \int_0^1{M_n^{+}(w; p, \mathbf{a})dp},$$
$$\mu_1^{-}(w) = \int_0^1{M_n^{-}(w; p, \mathbf{a})dp},$$

\begin{proof}
The upper partial moment:\\*
Considering that $M_n^{+}(w; p,\mathbf{a}) = 0\enspace\forall p < \alpha_w$ and $M_n^{+}(w; p,\mathbf{a}) = M_n(p,\mathbf{a})-w\enspace\forall p \geq \alpha_w$,
$$\int_0^1{M_n^{+}(w; p, \mathbf{a})dp} = \int_{\alpha_w}^1{M_n(p, \mathbf{a})dp} - \int_{\alpha_w}^1{wdp} = (1-\alpha_w)\bar{q}_n(\alpha_w,X) - (1-\alpha_w)w = \mu_1^{+}(w).$$

The lower partial moment:\\*
Considering that $M_n^{-}(w; p,\mathbf{a})=0\enspace\forall p > \alpha_w$ and $M_n^{-}(w; p,\mathbf{a}) = w - M_n(p,\mathbf{a})\enspace\forall p \leq \alpha_0$,
\begin{align*}
    & \int_0^1{M_n^{-}(w; p,\mathbf{a})dp} = \int_0^{\alpha_0}{wdp} - \int_0^{\alpha_0}{M_n(p, \mathbf{a})dp} = w\alpha_w - \left(\int_0^1{M_n(p, \mathbf{a})dp}-\int_{\alpha_0}^1{M_n(p, \mathbf{a})dp}\right) =\\
    &= w\alpha_w - E[X] + (1-\alpha_w)\bar{q}_n(\alpha_w,X) = \mu_1^{-}(w).
\end{align*}

\end{proof}

\begin{cor} \label{cor:part_mom3}
The upper and the lower partial moments at w for any metalog distribution are convex with respect to $\mathbf{a}$.

\begin{proof}
The metalog quantile function is linear with respect to $\mathbf{a}$, so $\forall i,j$ $\frac{\partial^2 M_n(p, \mathbf{a})}{\partial a_i^2} = 0$ and $\frac{\partial^2 M_n(p, \mathbf{a})}{\partial a_i\partial a_j} = 0$, so the Hessian matrix is zero and thus it is positive semi-definite, which implies $M_n(p, \mathbf{a})$ is convex with respect to $\mathbf{a}$. The same is actually true for $-M_n(p, \mathbf{a})$, as its Hessian matrix is also zero and thus positive semi-definite. Being a maximum of two convex functions, both $M_n^{+}(w; p, \mathbf{a})$ and $M_n^{-}(w; p, \mathbf{a})$ are convex with respect to $\mathbf{a}$. As an integral of a convex function is also convex if it exists, $\mu_1^{+}(w)$ and $\mu_1^{-}(w)$ are convex with respect to $\mathbf{a}$.
\end{proof}

\end{cor}

\end{prop}

\end{document}